\documentclass[letterpaper, 10pt, conference]{ieeeconf}      % Use this line for a4 paper

\IEEEoverridecommandlockouts                              % This command is only needed if 
\usepackage{graphicx}
\usepackage{color}
\usepackage[cmex10]{amsmath}
\usepackage{amssymb}
\usepackage[tight,footnotesize]{subfigure}
\usepackage{float}
\usepackage{multirow}
\usepackage{booktabs}
\usepackage{subeqnarray}
\usepackage{enumerate}
\usepackage{epstopdf}
\usepackage{wrapfig}
\usepackage{soul}
\usepackage[table]{xcolor}
\usepackage{array}
\usepackage{algorithm}
\usepackage{algpseudocode}
\usepackage{algorithmicx}
\usepackage{ctable}

\usepackage{cite}
\usepackage[bookmarks=true,colorlinks=true,linkcolor=black,backref=true,
pdfborder={0 0 0},citecolor=blue,urlcolor=blue]{hyperref}

\makeatletter
\let\proof\@undefined			% undefine \proof
\let\endproof\@undefined		% undefine \endproof
\makeatother
\usepackage{amsthm}

\theoremstyle{definition}

\newcommand{\real}[1][]{\mathbb{R}^{#1}}                                % Set of reals
\newcommand{\integer}[1][]{\mathbb{Z}^{#1}}                             % Set of integers
\newcommand{\defeq}{:=}                                                 % Definition 
\newcommand{\msub}[1]{_\mathrm{#1}}                                     % Roman subscript
\newcommand{\msup}[1]{^\mathrm{#1}}                                     % Roman superscript

\newcommand{\inv}[1]{{#1}^{-1}}                                         % Inverse (matrix)
\newcommand{\half}{ {\textstyle{\frac{1}{2}}} }							% Half
\newcommand{\clint}[2]{\left[#1, #2\right]}                             % Closed interval
\renewcommand{\geq}{\geqslant}                                          % Slanted geq

\newcommand{\expect}[1]{\mathbb{E}\left[#1\right]}

\newcommand{\mydef}[1]{{\textit{#1}}}

\newcommand{\ppl}{ath-planning}

\newcommand{\eqnnt}[1]{\hyperref[#1]{(\ref*{#1})}}
\newcommand{\eqnsnt}[2]{\hyperref[#1]{(\ref*{#1})}
	and~\hyperref[#2]{(\ref*{#2})}}
\newcommand{\eqnsernt}[2]{\hyperref[#1]{(\ref*{#1})}--\hyperref[#2]{(\ref*{#2})}}

\newcommand{\eqn}[1]{\hyperref[#1]{Eqn.~(\ref*{#1})}}
\newcommand{\eqns}[2]{\hyperref[#1]{Eqns.~(\ref*{#1})} and~\hyperref[#2]{(\ref*{#2})}}
\newcommand{\eqnser}[2]{\hyperref[#1]{Eqns.~(\ref*{#1})}--\hyperref[#2]{(\ref*{#2})}}
\newcommand{\eqnf}[1]{\hyperref[#1]{Equation~(\ref*{#1})}}
\newcommand{\eqnfs}[2]{\hyperref[#1]{Equations~(\ref*{#1})} and~\hyperref[#2]{(\ref*{#2})}}

\newcommand{\scn}[1]{\hyperref[#1]{Section~\ref*{#1}}}
\newcommand{\scns}[2]{\hyperref[#1]{Sections~\ref*{#1}} and~\hyperref[#2]{\ref*{#2}}}
\newcommand{\scnser}[2]{\hyperref[#1]{Sections~\ref*{#1}}--\hyperref[#2]{\ref*{#2}}}

\newcommand{\fig}[1]{\hyperref[#1]{Fig.~\ref*{#1}}}
\newcommand{\figs}[2]{\hyperref[#1]{Figs.~\ref*{#1}} and~\hyperref[#2]{\ref*{#2}}}
\newcommand{\figser}[2]{\hyperref[#1]{Figs.~\ref*{#1}}--\hyperref[#2]{\ref*{#2}}}
\newcommand{\figf}[1]{\hyperref[#1]{Figure~\ref*{#1}}}
\newcommand{\figfs}[2]{\hyperref[#1]{Figures~\ref*{#1}} and~\hyperref[#2]{\ref*{#2}}}
\newcommand{\figfser}[2]{\hyperref[#1]{Figures~\ref*{#1}}--\hyperref[#2]{\ref*{#2}}}

\newcommand{\tbl}[1]{\hyperref[#1]{Table~\ref*{#1}}}
\newcommand{\tbls}[2]{\hyperref[#1]{Tables~\ref*{#1}} and~\hyperref[#2]{\ref*{#2}}}

\newcommand{\apx}[1]{\hyperref[#1]{Appendix~\ref*{#1}}}

\newcommand{\exmpl}[1]{\hyperref[#1]{Example~\ref*{#1}}}
\newcommand{\exmpls}[2]{\hyperref[#1]{Examples~\ref*{#1}} and~\hyperref[#2]{\ref*{#2}}}
\newcommand{\exmplser}[2]{\hyperref[#1]{Examples~\ref*{#1}}--\hyperref[#2]{(\ref*{#2})}}

\newcommand{\chp}[1]{\hyperref[#1]{Chapter~\ref*{#1}}}
\newcommand{\chps}[2]{\hyperref[#1]{Chapters~\ref*{#1}} and~\hyperref[#2]{(\ref*{#2})}}
\newcommand{\chpser}[2]{\hyperref[#1]{Chapters~\ref*{#1}}--\hyperref[#2]{(\ref*{#2})}}

\newcommand{\prb}[1]{\hyperref[#1]{Problem~\ref*{#1}}}
\newcommand{\prp}[1]{\hyperref[#1]{Prop.~\ref*{#1}}}
\newcommand{\prpf}[1]{\hyperref[#1]{Proposition~\ref*{#1}}}

\newcommand{\thmref}[1]{\hyperref[#1]{Theorem~\ref*{#1}}}
\newcommand{\thmsref}[2]{\hyperref[#1]{Theorems~\ref*{#1}} and~\hyperref[#2]{\ref*{#2}}}
\newcommand{\thmserref}[2]{\hyperref[#1]{Theorems~\ref*{#1}}--\hyperref[#2]{\ref*{#2}}}

\newcommand{\algline}[1]{\hyperref[#1]{Line~\ref*{#1}}}
\newcommand{\alglines}[2]{\hyperref[#1]{Lines~\ref*{#1}} and~\hyperref[#2]{\ref*{#2}}}
\newcommand{\alglineser}[2]{\hyperref[#1]{Lines~\ref*{#1}}--\hyperref[#2]{\ref*{#2}}}

\renewcommand{\vec}[1]{\boldsymbol{#1}}

\def\graph{\mathcal{G}}
\def\verts{V}
\def\edges{E}

\def\nVertsX{N_1}
\def\nVertsY{N_2}

\newcommand{\edgeDistance}[2]{d_{#1#2}}
\newcommand{\edgeCost}[2]{c_{#1#2}}
\newcommand{\windTrue}[2]{w_{#1#2}}
\newcommand{\windMeas}[2]{\tilde{w}_{#1#2}}
\newcommand{\windFit}[2]{\hat{w}_{#1#2}}
\def\pressure{P}
\newcommand{\dpx}[1]{{\pressure}_{x_{#1}}}
\newcommand{\edpx}[1]{{\hat{\pressure}}_{x_{#1}}}

\newcommand{\edgeResist}[2]{\varrho_{#1#2}}
\newcommand{\airspeed}{u_0}
\newcommand{\agentWind}{w}
\newcommand{\timeStep}{\Delta t}
\newcommand{\scaleFactor}{r}

\title{\LARGE \bf
Minimum-Time Sequential Traversal by a Team of Small Unmanned Aerial Vehicles in an Unknown Environment with 
Winds
}

\author{Jeffrey A. DesRoches$^{1}$ and Raghvendra V. Cowlagi$^{1}$% <-this % stops a space
\thanks{$^{1}$Jeffrey A. DesRoches and Raghvendra V. Cowlagi are with the Department of Aerospace Engineering, 
Worcester Polytechnic Institute, Worcester, MA 01609.
        Email: {\tt\footnotesize jadesroches, rvcowlagi@wpi.edu}}%
}

\begin{document}

\maketitle
\thispagestyle{empty}
\pagestyle{empty}

%%%%%%%%%%%%%%%%%%%%%%%%%%%%%%%%%%%%%%%%%%%%%%%%%%%%%%%%%%%%%%%%%%%%%%%%%%%%%%%%
\begin{abstract}
	
	We consider the problem of transporting multiple packages 
	from an initial location to a destination location in
	a windy urban environment using a team of SUAVs. Each SUAV 
	carries one package.
	We assume that the wind field is unknown, but wind speed can be measured
	by SUAVs during flight. The SUAVs fly sequentially one after the other, 
	measure wind speeds along their trajectories, and report the measurements 
	to a central computer.
	The overall objective is to minimize the total travel time of all SUAVs, 
	which is in turn related to the number of SUAV traversals through
	the environment.
	For a discretized environment modeled by a graph, we describe 
	a method to estimate wind speeds and the time of traversal for each
	SUAV path. Each SUAV traverses a minimum-time path planned
	based on the current wind field estimate.
	We study cases of static and time-varying wind fields
	with and without measurement noise. For each case, we demonstrate
	via numerical simulation that the proposed method finds the optimal
	path after a minimal number of traversals.

\end{abstract}

%%%%%%%%%%%%%%%%%%%%%%%%%%%%%%%%%%%%%%%%%%%%%%%%%%%%%%%%%%%%%%%%%%%%%%%%%%%%%%%%
\section{Introduction}

Urban air mobility promises to revolutionize the transportation of passengers
and goods in densely populated areas~\cite{Cohen2021}, with envisioned applications
in time-sensitive package deliveries~\cite{Zhao2024} and emergency response
services~\cite{goyal2018urban}.

A key challenge in urban air mobility, and especially for small uncrewed aerial vehicles
(SUAVs) designed for package delivery, to navigate urban environments is the 
presence of dynamic wind conditions. Due to the ``urban canyon''
effect, wind gusts at relatively low altitudes in urban areas can reach
speeds as high as $20$ m/s~\cite{Rakib2020}, which is comparable to --
or at least a significant fraction of -- the cruise speeds of many SUAVs.
Low-altitude urban wind fields are complex and difficult to model, 
with relatively few datasets and studies of reconstructing the wind field from data, 
e.g.,~\cite{CHRIT2023100451}. 
Some recent studies report on SUAV p\ppl\ in urban areas using wind data reconstructed
with computational fluid dynamics in \cite{doi:10.2514/6.2020-2890, WU2024105253}. 
However, these are limited to static fields that are assumed to be fully known 
to the p\ppl\ algorithm.

The problem of SUAV p\ppl\ in dynamic urban wind fields is challenging 
not merely due to the lack of pre-existing datasets. 
Even if the typical wind fields in a particular
urban area were easily available, these data would not be helpful of the wind fields
prevalent during unusual weather patterns or relatively rare natural events
such as hurricanes. Furthermore, existing datasets examine the wind field at coarse
temporal resolutions (e.g., 10 minutes) because the timescales 
considered are typically seasonal or annual~\cite{refId0}.

In this paper, we consider the problem of transporting multiple
packages from an initial location to a destination location in
a windy urban environment using a team of SUAVs. Each SUAV 
carries one package. The overall objective is to minimize the
total travel time of all SUAVs.
We assume that the wind field is unknown, but wind speed can be measured
by SUAVs during flight along their trajectories. 
The SUAVs fly one after the other, measure wind speeds along their trajectories, 
and report the measurements to a central computer,
henceforth referred to as the \mydef{agent.} 
The flight of a single SUAV in the team from the 
initial to the goal location is called a \mydef{pass.}
The central computer calculates an estimate of the wind field
based on measurements collected in each pass and
transmits the estimate to all other SUAVs.

A na\"ive solution to this problem is to make the first one or two SUAVs
follow ``lawnmower'' patterns to fully survey the entire environment.
This would enable the remaining SUAVs to plan paths using accurate and
precise estimates of the wind field. However, this solution will necessitate
a large time of travel by the initial surveying SUAVs. Furthermore, this 
solution may not be implementable in large environments, where a single SUAV
may not have enough range and endurance to survey the environment.
Therefore, we seek a solution for planning the path of each SUAV
towards the overall objective of minimizing the team's total travel time.

Typically, the goal of p\ppl\ is to find a path that minimizes some cost function. 
The most commonly used p\ppl\ methods include cell- or grid-based disretization, 
probabilistic roadmaps, and artificial potential field techniques
\cite{LaValle2006,Patle2019}. In cell- or grid-based discretization methods,
Dijkstra's algorithm or $\textrm{A}^{\ast}$ are employed to efficiently 
compute the shortest path. Other emerging approaches, such as deep reinforcement 
learning \cite{wen2024drl, qin2023deep} claim to handle complex and uncertain, 
yet static environments. However, these methods typically large volumes of training data
obtained from historical datasets or simulations. 
This study aims to remove this requirement altogether. Our assumption is that all 
relevant data regarding the wind field is to be collected during the aforesaid 
SUAV passes.

Measurements from sensors are noisy and may be intermittently unavailable. 
Typical estimation techniques such as the Kalman filter~\cite{Lewis2017optimal}, 
maximum likelihood estimator~\cite{myung2003tutorial}, and Bayesian filter~\cite{thrun2006} 
are all well studied estimators for linear systems. The extended Kalman filter, 
unscented Kalman filter \cite{Julier2004} and the particle filter \cite{deng2021poserbpf} 
are common for nonlinear dynamical systems. The common underlying assumption for these
estimators is the presence of a model of the system's evolution as well as a measurement
model. If the system model is unknown or highly imprecise (i.e., large process noise covariance),
then these estimators effectively perform least squares regression on the measured data.
For unmodeled systems with noisy measurements,
other system identification least squares methods~\cite{HEIJ1999993} are reported. 
For periodically time-varying systems, frequency domain total least squares are proposed in 
\cite{LOUARROUDI201113115,LIU2020736}.

Other related work for sensor configuration in unknown environments is 
explored by \cite{poudel2024coupled, Poudel-Cowlagi-Scitech2025}, which is 
based on an iterative process that identifies the optimal sensor 
configuration. However, these studies utilize a separate ego vehicle from 
the ``sensor vehicles.'' This approach is more appropriate for scenarios 
with a heterogeneous mix of vehicles such as a wheeled ego vehicle and UAV 
sensor vehicles.

The literature on  wind estimation onboard UAVs largely addresses path- or 
trajectory tracking in the presence of wind 
disturbances~\cite{sarras2014guidance,Zhou2017,Yang2021}, 
gust rejection for stability in cruise or hover 
conditions~\cite{Xing2023,liu2016disturbance},
and/or dynamical modeling~\cite{Wang2019}. Wind estimation is studied
in the airdrop literature~\cite{Cacan2015} and in the study of 
weather systems~\cite{cecil2017hurricane}.

The novelty of this work in relation to the literature is that we consider
the wind estimation and package delivery problems simultaneously. The SUAVs 
are constrained to travel between a fixed initial point and a fixed destination,
and do not travel any other special paths for estimating the wind. We explore 
strategies for reducing the total travel time of a team of 
SUAVs in unknown wind without modifying the p\ppl\ algorithm from a typical
minimum-cost approach. Therefore, the proposed method can be easily implemented
with off-the-shelf SUAVs with little to no modification.

{The rest of this paper is organized as follows.} In \scn{sec-problem} we 
introduce the main elements of the scenario(s) considered in this paper.
In \scn{sec-method}, we describe the proposed method for estimating the
unknown wind via data collected in SUAV passes. In \scn{sec-results}, we
provide numerical simulation results, and conclude the paper in \scn{sec-conclusions}.

%%%%%%%%%%%%%%%%%%%%%%%%%%%%%%%%%%%%%%%%%%%%%%%%%%%%%%%%%%%%%%%%%%%%%%%%%%%%%%%%%%%%%
\section{Problem Formulation}

\label{sec-problem}

\begin{figure}
	\centering
	\includegraphics[width=\columnwidth]{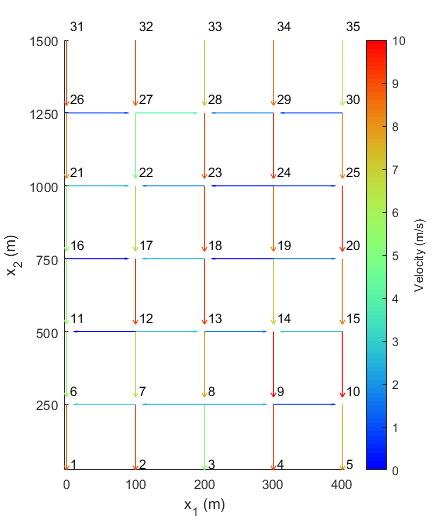}
	\caption{Illustrative example of a wind field model on a
		$5 \times 5$ grid. Note that the top and bottom rows are only 
		used as boundary conditions.}
	\label{fig:velocitymap}
\end{figure}

Consider a 2D environment modeled by a {directed} graph $\graph = (\verts,\edges).$
The vertex set $\verts$ consists of uniformly spaced grid points in the environment.
Let $\nVertsX$ and $\nVertsY$ denote the number of grid points
along the $x_1$ and $x_2$ principal axes, respectively, in some prespecified Cartesian system.
The edge set $\edges$ consists of geometrically adjacent vertices as illustrated
in~\fig{fig:velocitymap}. Adjacency is defined in the ``4-neighbor'' sense, i.e., 
grid points to the top, down, left, and right are considered adjacent, but diagonal adjacency
is not considered. This assumption reflects the typical road network in an urban area.
The vertices are labeled by integers, i.e., $\verts = \{1, 2,..., \nVertsX\nVertsY\}.$

The distance along an edge $(i,j) \in \edges$ with $i,j \in \verts$ 
is denoted $\edgeDistance{i}{j}.$ The time-varying wind speed along edge $(i,j)$
is denoted $\windTrue{i}{j}(t),$ and it is assumed to be bounded $\windTrue{i}{j}(t)
\in \clint{-w\msub{max}}{w\msub{max}}$ and anti-symmetric
$\windTrue{i}{j}(t) = -\windTrue{j}{i}(t)$ for all time $t \in \real_{\geq 0}.$
The wind speed is assumed to be independent of the spatial variables $x_1$ and $x_2.$
The true values of $\windTrue{i}{j}(t)$ are unknown to the SUAV team.

To estimate the wind speeds based on measurements, an underlying physics-based model
is necessary. A high-fidelity fluid dynamical model is impractical for use in
estimation, and therefore we postulate a simpler model, 
analogous to a simple electrical circuit. The wind speeds on each edge are assumed 
to be proportional to the spatial gradient of some 
(unknown) spatiotemporal {``pressure''} field denoted $\pressure(x_1, x_2, t),$
analogous to a potential difference in electrical circuits.
In this paper we assume $\dpx{1}(t) \defeq \partial \pressure/\partial x_{1}(t) = 0,$
i.e., the only non-zero spatial gradients of the pressure field 
is $\dpx{2}(t) \defeq \partial \pressure/\partial x_{2}(t).$
In this paper, we focus on constant or time-periodic signals $\dpx{2}(t).$ 
The dependence of the wind speeds on this pressure gradient is as follows.

Edges along the $x_{2}$ direction are assumed to have fixed but unknown and
strictly positive {``resistance''} values. Edges along the $x_1$ direction
are assumed to have zero resistance values, due to which the edges along
$x_2$ may be treated as parallel resistances in a circuit.
These assumptions reflect typical Manhattan-type urban layouts with
wide avenues in one direction and narrower roads connecting the wide 
avenues. With these assumptions the wind speeds on each edge can be 
easily calculated based on conservation of mass flow from one boundary of
the environment to the other. Due to this model, we can 
transform the wind speed estimation problem as into the estimation
of the signal $\dpx{2}(t).$ Specifically, due to this model, it is easy
to show that the wind speed on any edge $(i,j) \in \edges$ is proportional
to the signal $\dpx{2}(t).$ We denote the constants of proportionality,
which are unknown to the SUAVs, by $\edgeResist{i}{j}.$

\begin{figure}
	\centering
	\includegraphics[width=\columnwidth]{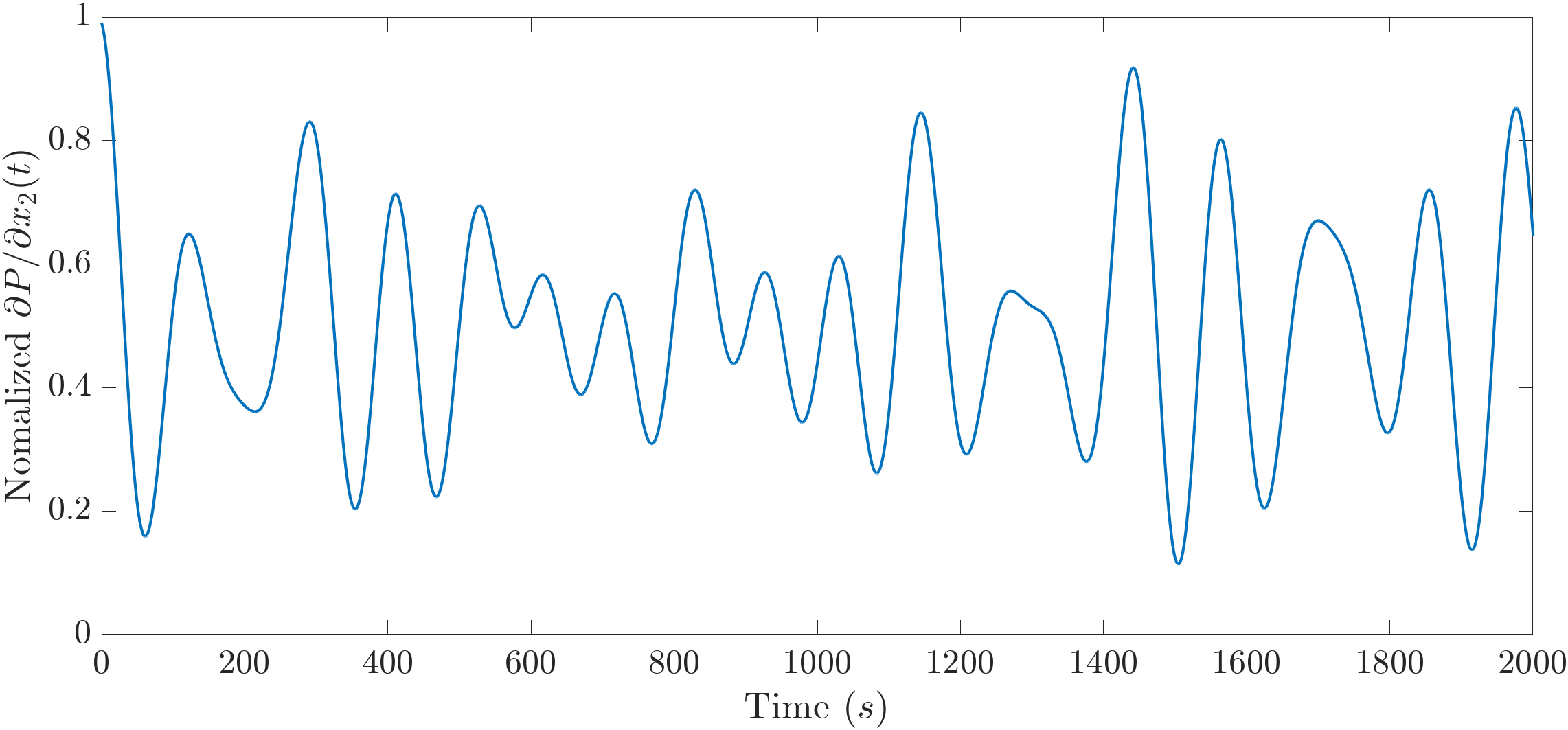}
	\caption{An example of a periodic signal representing
		$\dpx{2}(t).$}
	\label{fig:dpdx2noise}
\end{figure}

An example wind field generated with this method is shown in 
\fig{fig:velocitymap}. The first and last rows are used as 
nontraversable boundaries necessary to construct the field. 
In \fig{fig:velocitymap}, the start is assumed to be the bottom 
left corner (vertex 6) and the goal is the top right corner (vertex 30), 
since we must omit the top and bottom rows of vertices.
An example of the $\dpx{2}(t)$ considered is illustrated in
\fig{fig:dpdx2noise}, which is obtained by a weighted finite 
sum of sine and cosine signals 
(details in \scn{sec-results}).
%The objective of this process is to create a simple yet diverse graph which 
%adheres to conservation of mass and contains a unique, globally optimal path between the start and 
%goal 
%vertices. 

We assume that the measurements that each SUAV can make are
of $\windTrue{i}{j}$ with additive measurement noise for the 
edges along which the SUAV travels. The measured wind speed
is denoted by $\windMeas{i}{j}$ to distinguish from the true 
wind speed $\windTrue{i}{j}.$ Each SUAV
travels at a fixed air-relative speed denoted $\airspeed.$

%%%%%%%%%%%%%%%%%%%%%%%%%%%%%%%%%%%%%%%%%%%%%%%%%%%%%%%%%%%%%%%%%%%%%%%%%%%%%%%%%%%%%
\section{Method}
\label{sec-method}

In what follows, $\integer$ and $\real$ are the sets of integers and real numbers,
respectively. $\expect{\cdot}$ denotes the expectation operator.

We consider a sequence of SUAV passes from the initial location to the destination.
As previously mentioned, all data is stored and processed by a central agent. We
neglect any issues related to the communication of data from the SUAVs to the agent.
The agent's computations are in discrete time, which we denote by integers
$k, m, n,$ etc. to indicate time instants $t_k, t_m, t_n,$ etc. Successive
discrete time instants are all separated by a uniform and fixed time step of $\timeStep.$
The sequence of all time instants in succession in an interval from $n$ to $m$
with $n, m \in \integer$ and $n < m$ is denoted by $n:m.$

The agent maintains an estimate of the transition cost, 
denoted $\edgeCost{i}{j},$ for each edge $(i,j) \in \edges.$ 
Because we are interested in minimum-time travel, the transition cost
of an edge is equal to the time taken to travel along that edge.
Before the first pass, when no wind speed data are available, the cost
estimates for all edges are uniformly initialized to 
\begin{align}
	\edgeCost{i}{j} &= {\edgeDistance{i}{j}} \inv{(\airspeed + w\msub{max})},
	\quad \mbox{for each } (i,j) \in \edges.
	\label{eq-cost-init}
\end{align}
This initialization is ``optimistic'' in that it assumes the highest possible
tailwind along each edge, and is therefore a lower bound on the true edge 
cost. Note that this optimistic initialization disregards the anti-symmetric
property of wind speeds on edges.

For each pass, the agent plans the minimum-time path for the SUAV. This is
easily achieved by executing Dijkstra's algorithm on the graph $\graph$ using
the most recent estimates of the costs $\edgeCost{i}{j}.$ Due to the optimistic
initializiation in \eqnnt{eq-cost-init}, in the first few passes
there may be several paths with equal cost, of which one is randomly chosen. 
This may be accomplished within Dijkstra's
algorithm itself at the crucial step of popping the fringe (also known
as the ``open'' list), i.e., choosing the vertex-label pair with minimum label.

The SUAV then traverses this planned path, i.e., executes the pass. Consider
an edge $(i,j) \in \edges$ traversed within this pass, and denote by $n, m \in \integer$
the discrete time instants at which the SUAV enters and exits this edge, respectively.
During this traversal, the SUAV measures and provides to the agent
the wind speed time series $\windMeas{i}{j}(n:m).$
The agent's objective then is to use these measurements to update
the edge cost estimates.

To this end, we consider four different cases as follows:
\begin{enumerate}[{Case}~1:]
	\item The true wind speed $\windTrue{i}{j}$ is time-invariant
	for each $(i,j) \in \edges$ and there is no measurement noise, i.e.,
	$\windMeas{i}{j} = \windTrue{i}{j}.$
	\item The true wind speed $\windTrue{i}{j}$ is time-invariant
	for each $(i,j) \in \edges$ and there is zero-mean measurement noise, i.e.,
	$\expect{\windMeas{i}{j} - \windTrue{i}{j}} = 0.$
	\item The true wind speed $\windTrue{i}{j}$ is time-varying
	for each $(i,j) \in \edges$ and there is no measurement noise, i.e.,
	$\windMeas{i}{j}(k) = \windTrue{i}{j}(k)$ for each $k \in \integer.$
	\item The true wind speed $\windTrue{i}{j}$ is time-invariant
	for each $(i,j) \in \edges$ and there is zero-mean measurement noise, i.e.,
	$\expect{\windMeas{i}{j}(k) - \windTrue{i}{j}(k)} = 0$
	for each $k \in \integer.$
\end{enumerate}

%%%%%%%%%%%%%%%%%%%%%%%%%%%%%%%%%%%%%%%%%%%%%%%%%%%%%%%%%%%%%%%%%%%%%%%%%%%%%%%%%%
\subsection{Edge Cost Estimates}

The agent's method of updating the edge cost estimates differs for 
each of the aforesaid Cases. For Case~1, the edge costs may be updated simply
by considering the SUAV's time of travel along the edge $(i,j)$ in
relation to the (known) distance $\edgeDistance{i}{j}$ of travel.
Specifically, the agent updates $\edgeCost{i}{j}$ as
\begin{align}
	\edgeCost{i}{j} &= { \edgeDistance{i}{j} }
	\inv{(\airspeed + \windMeas{i}{j})}. 
	\label{eq:cost-case1}
\end{align}

For Case~2, where zero mean measurement noise is present, the agent updates
the edge costs by considering the mean value of the measured wind speeds, i.e.,
\begin{align}
	\edgeCost{i}{j} &= { \edgeDistance{i}{j} }
	\inv{\left(\airspeed + \frac{1}{m-n}\sum_{k=n}^{m}\windMeas{i}{j}(k)\right)}. 
	\label{eq:cost-case2}
\end{align}

\begin{figure}
	\centering
	\includegraphics[width=\columnwidth]{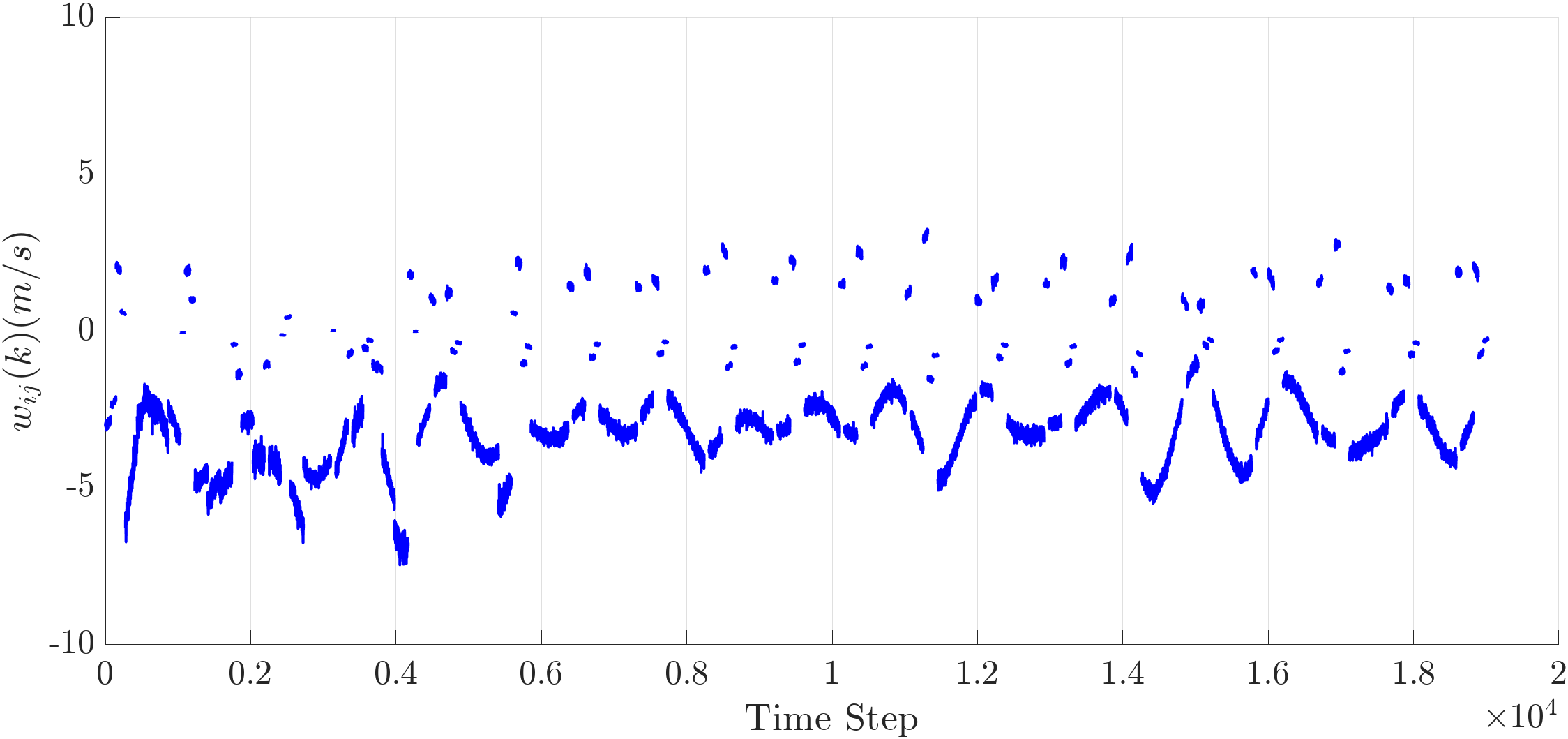}
	\caption{Raw $\windMeas{i}{j}(k)$ measurement data for a simulated 
		Case~3 on the $5\times 5$ grid shown in \fig{fig:velocitymap}.}
	\label{fig:edgevelocityclean}
\end{figure}

For Cases~3 and~4, where the wind speeds are time-varying, 
the following crucial problem arises. The 
observations of the wind speeds are restricted to the time
window $n:m$ during which an SUAV traverses the edge. Furthermore,
because the edge ``resistance'' values are different, when the SUAV
transitions from one edge to the next, there is in general a 
discontinuity in the measured wind speed. For example, suppose
$(i,j)$ and $(j,\ell)$ are successive edges in the SUAV's path.
The SUAV completes traveling along $(i,j)$ at time $m$
and starts to traverse edge $(j,\ell)$ at time $m+1.$ In general
there will be a discontinuity such that $| \windMeas{i}{j}(m) - \windMeas{j}{\ell}(m+1) |$
cannot be guaranteed to be smaller than some $\varepsilon$ dependent
on the time step $\timeStep.$

\figf{fig:edgevelocityclean} illustrates
the (noiseless) measurements of wind speeds along edges a numerical simulation
of Case~3 and these discontinuities are clearly visible.
The presence of these discontinuities makes the time-series estimation
of wind speeds and, consequently, the edge costs, nontrivial. 
To resolve this issue, we focus instead on the estimation of the
time-series $\dpx{2}(k),$ which \emph{is} continuous.

Based on the model postulated in \scn{sec-problem}, recall that the
wind speed along each edge is proportional to the pressure gradient
$\dpx{2}(k)$:
\begin{align}
	\windTrue{i}{j}(n:m) &= \edgeResist{i}{j} \dpx{2}(n:m).
	\label{eq-wind-resistance}
\end{align}
Let $\edpx{2}(n:m)$ denote the mean value of an estimate of $\dpx{2}(n:m).$
We may then approximate the unknown constant $\edgeResist{i}{j}$ using
the time-averaged values of the measured wind speed and of the estimate $\edpx{2}$ as:
\begin{align}
	\edgeResist{i}{j} &= \frac{1}{m-n} 
	{\left(\sum_{k=n}^{m}\windMeas{i}{j}(n:m)\right)} 
	\inv{\left(\sum_{k=n}^{m}\edpx{2}(n:m)\right)}.
	\label{eq-resistance-calc}
\end{align}
The agent can then approximate the edge cost for edge $(i,j)$ by an expression
similar to~\eqnnt{eq:cost-case2}:
\begin{align}
	\edgeCost{i}{j} &= { \edgeDistance{i}{j} }
	\inv{\left(\airspeed + \edgeResist{i}{j}\right)}. 
	\label{eq:cost-case2}
\end{align}
Note that this is a static (time-invariant) approximation of the edge cost,
which in principle will be time-varying when the wind speeds are time-varying.
Searching a graph with time-varying edge costs can significantly worsen
computational complexity.

Based on~\eqnsernt{eq-wind-resistance}{eq-resistance-calc}, 
we focus on the problem of finding the estimate $\dpx{2}(k)$ 
in Cases~3 and~4, namely, where measurement noise is either absent or 
is of zero mean.

%%%%%%%%%%%%%%%%%%%%%%%%%%%%%%%%%%%%%%%%%%%%%%%%%%%%%%%%%%%%%%%%%%%%%%%%%%%%%%%%%%
\subsection{$\dpx{2}(k)$ Estimation}

For Cases~3 and~4, we consider first a simple least-squares
polynomial estimate for $\dpx{2}(k).$ 
To this end, the agent maintains a time series $\agentWind: \integer \rightarrow \real$
that ``stitches'' the time-window measurements $\windMeas{i}{j}(n:m)$ into a single 
time series that extends over all SUAV passes. This stitching is achieved
by scaling the values of the new window of measurements to the larger time series,
followed by concatenation. The desired signal $\edpx{2}(k)$ is then recovered by 
normalizing $\agentWind.$ Specifically, we concatenate $\agentWind(1:n-1)$ with the new
measurements by
\begin{align*}
	\agentWind(n:m) \defeq \scaleFactor\windMeas{i}{j}(n:m),
\end{align*}
where $\scaleFactor \defeq \inv{(\windMeas{i}{j}(n))} (\agentWind(n-1) + \dot{\agentWind}(n-1) 
\timeStep)$ is a scaling factor. The value of the derivative $\dot{\agentWind}(n-1)$
is calculated numerically using a second-order backward difference approximation.
Finally, we normalize the concatenated time series to find
\begin{align*}
	\edpx{2}(1:m) \defeq \frac{\agentWind(1:m)}{\max_{k \in [1,m]}|\agentWind(k)|}.
\end{align*}

Whereas this method suffices for Case~3 (no measurement noise), the numerical
derivative calculation obviously fails for Case~4. In this Case, we may 
use a polynomial approximation of $\windMeas{i}{j}(n:m).$ To avoid overfitting,
we consider a quadratic polynomial. Let $\windFit{i}{j}(k)$ denote a quadratic
least-squares approximation to the measured signal $\windMeas{i}{j}(n:m).$
Then, we redefine the aforesaid scale factor as
\begin{align*}
	\scaleFactor \defeq \inv{(\windFit{i}{j}(n-\half))} \agentWind(n-\half),
\end{align*}
while the rest of the calculations remain similar.

The advantage of the method discussed so far is that we make 
no assumptions about the nature of the signal $\dpx{2}(t)$ other
than it is once-differentiable with respect to time.
The shortcoming of this method is that, in the presence of measurement noise,
the error propagates and grows over time. Therefore, we develop a Kalman filter,
while noting a restriction of its applicability to a specific class of signals.

Consider a state variable of the form $$\vec{x} = (y_1, y_2, \ldots, y_N,
\dot{y}_1, \dot{y}_2, \ldots, \dot{y}_N, \ddot{y}_1, \ddot{y}_2, \ldots, \ddot{y}_N),$$
where $y_N(t) = \dpx{2}(t)$ and $y_1,\ldots, y_{N-1}$ are hidden variables.
We write a state evolution model and observation model of the form
\begin{align}
	\vec{x}(k) &= A\vec{x}(k-1) + \vec{\eta}_1(k-1), \\
	\vec{z}(k) &= C\vec{x}(k) + \vec{\eta}_2(k),
	\label{eq:systemmodel}
\end{align}
where $\vec{\eta}_1$ and $\vec{\eta}_2$ are zero mean Gaussian white
noise processes with error covariances $Q$ and $R,$ respectively.

The hidden variables may be thought of as displacements of a series of $N$
fictitious mass-spring systems. Without damping, we focus on the modeling
of periodic signals that are exactly expressed by an $N-$term Fourier series
expansion.

Stated differently, we may perform a fast Fourier transform (FFT)
analysis on a historical dataset of $\dpx{2}(k)$ or on the data collected
during passes to find the $N$ most dominant frequency components, their 
amplitudes, and phase shifts.

In this model, we write the $A$ matrix as
\begin{equation}
	\begin{aligned}
		& A_{1:2N,1:3N} = \\
		&\begin{bmatrix}
			& 1 & 0 & \dots & \Delta t  & 0 & \dots & 0 \\
			& 0 & 1 & 0 & \dots & \Delta t  & 0 & 0 \\
			& \vdots & 	0 & 1 & 0 & \dots & \Delta t  & 0 \\
			& \vdots & \vdots & \ddots & \ddots & \ddots & \dots & \ddots \\
			& \end{bmatrix}, \\
		& A_{2M+1:3N,1:N} = \\
		& \begin{bmatrix}
			& -(K_1+K_2) & K_2 & 0 & \dots & 0  \\
			& -K_2 & -(K_2+K_3) & K_3 & 0 & 0   \\
			& 0 & \ddots & \ddots & \ddots & 0  \\
			& 0 & \dots & 0 & K_{N-1} & -K_N    \\
			& \end{bmatrix}, \\
		& A_{2N+1:3N,2N+1:3N} = 0,\\
		& C = \begin{bmatrix} 0 & \ldots & 1 & 0 & \ldots & 0 \end{bmatrix}.
	\end{aligned}	
\end{equation}
Here the constants $K_\ell$ are the fictitious spring stiffness values and
these are equal to the square roots of the frequencies resulting from the
FFT analysis mentioned above.

The amplitude and phase data extracted from the FFT can be used to set
a prior mean $\hat{\vec{x}}(0)$ and error covariance $\mathcal{P}(0)$
for the Kalman filter. Whereas the covariance $R$ may be determined 
from measurement error calibration, the covariance $Q$ must be tuned
based on the observed innovations (residual error) sequence. 
This is because the FFT analysis can provide inaccurate frequency
information with small time windows.
One strategy
for tuning~$Q$ is to set large diagonal values during the early passes
and then tuning it based on the FFT analysis after sufficient 
time has passed.

%%%%%%%%%%%%%%%%%%%%%%%%%%%%%%%%%%%%%%%%%%%%%%%%%%%%%%%%%%%%%%%%%%%%%%%%%%%%%%%%%%%%
\section{Results and Discussion}
\label{sec-results}

\subsection{Numerical Simulation Setup}

In this section, we report the results of numerical simulation studies of the proposed 
method. Inspired by the city blocks in Manhattan, (New York City, NY, USA) between the Hudson River 
and Central Park area, a directional graph model of the city roadways is produced, as 
illustrated in \fig{fig:velocitymap}. 
This model considers uniform road lengths of 100 and 250 meters along  the 
$x_{1}$ and $x_{2}$ directions, respectively. Winds speeds are calculated per the pressure gradient
model postulated in \scn{sec-problem} and all wind speeds are scaled such that 
the maximum speed is equal to a globally defined maximum of $w\msub{max} = 10$ m/s.
The air-relative speed of each SUAV is fixed at $\airspeed = 15$ m/s.

In the context of the pressure gradient model postulated in \scn{sec-problem},
all edge resistances are fixed at arbitrarily chosen values in $\clint{0.5}{1}.$
The ground truth pressure gradient $\dpx{2}(t)$ signal is produced via
the following expression including noise:
\begin{align}
	\dpx{2} (t) &= d_{0} + \sum_{\ell=1}^{N} a_{\ell} 
	\cos(2\pi b_{\ell}t + c_{\ell}) + v (t)
	\label{eq:fouriernoise}
\end{align}
where $a_{\ell}$, $b_{\ell}$, and $c_{\ell}$ are the amplitude, 
frequency, and phase respectively of each periodic 
component in the series, $v(t)$ simulates the measurement noise, 
and $d_{0}$ is a constant.

For repeated simulation trials, we need to randomly produce a ground truth
signal $\dpx{2}(t)$ with normalized values such that 
$\dpx{2}(t) \in \clint{0}{1}.$ To this end, $d_{0}$ is set to 0.5.
The amplitude coefficients $a_{\ell}$ are obtained by sampling 
a uniform distribution over $\clint{0}{1}$ and then normalized as 
\begin{equation*}
	a_{\ell} =\half {a_{\tilde{\ell}}} \inv{\textstyle{(\sum_{\tilde{\ell}=1}^{N} a_{\tilde{\ell}})}}.
	\label{eq:suma}
\end{equation*}
Similarly, the frequency coefficients $b_\ell$ are obtained by
sampling a uniform distribution over some finite frequency range of the user's choice.
The phase coefficients $c_\ell$ are sampled from a uniform distribution
over $\clint{0}{2\pi}.$

For Cases~3 and~4, $N=6$ terms were used to produce ground truths
$\dpx{2}(t)$ via \eqnnt{eq:fouriernoise}. The frequency terms were chosen
randomly between $300 \text{s}^{-1}$ and $500 \text{s}^{-1}.$
A sample $\dpx{2}$ is shown in \fig{fig:dpdx2noise}.
For Cases~2 and 4, the noise term was simulated as 
$v(k) \sim \mathcal{N}(0, 0.025).$

\subsection{Performance Analysis}

Because each pass is assumed to be conducted along a minimum-time path,
per the cost estimates available immediately before the pass,
the main performance metric of interest is the number of passes needed
until the optimal path is found.
We record also the time of travel for each pass as a measure
of performance, but this measure is dependent on various dimensional
assumptions, e.g., edge length, maximum wind speed, etc.

\begin{table}
	\caption{Sequential Trials until Optimal Path Convergence}
	\label{table_results}
	\begin{center}
		\begin{tabular}{c||c|c|c|c}
			\specialrule{.3em}{.2em}{.2em}
			& Case 1 & Case 2 & Case 3 & Case 4\\
			\specialrule{.2em}{.1em}{.1em}
			5x5 & 6 & 6 & 6 & 7   \\ 
			\hline
			7x7 & 12 & 12 & 12& 14 \\ 
			\hline
			9x9 & 15 & 16 & 15& 18  \\ 
			\specialrule{.3em}{.2em}{.2em}
		\end{tabular}
	\end{center}
\end{table}

\begin{figure}
	\centering
	\includegraphics[width=\columnwidth]{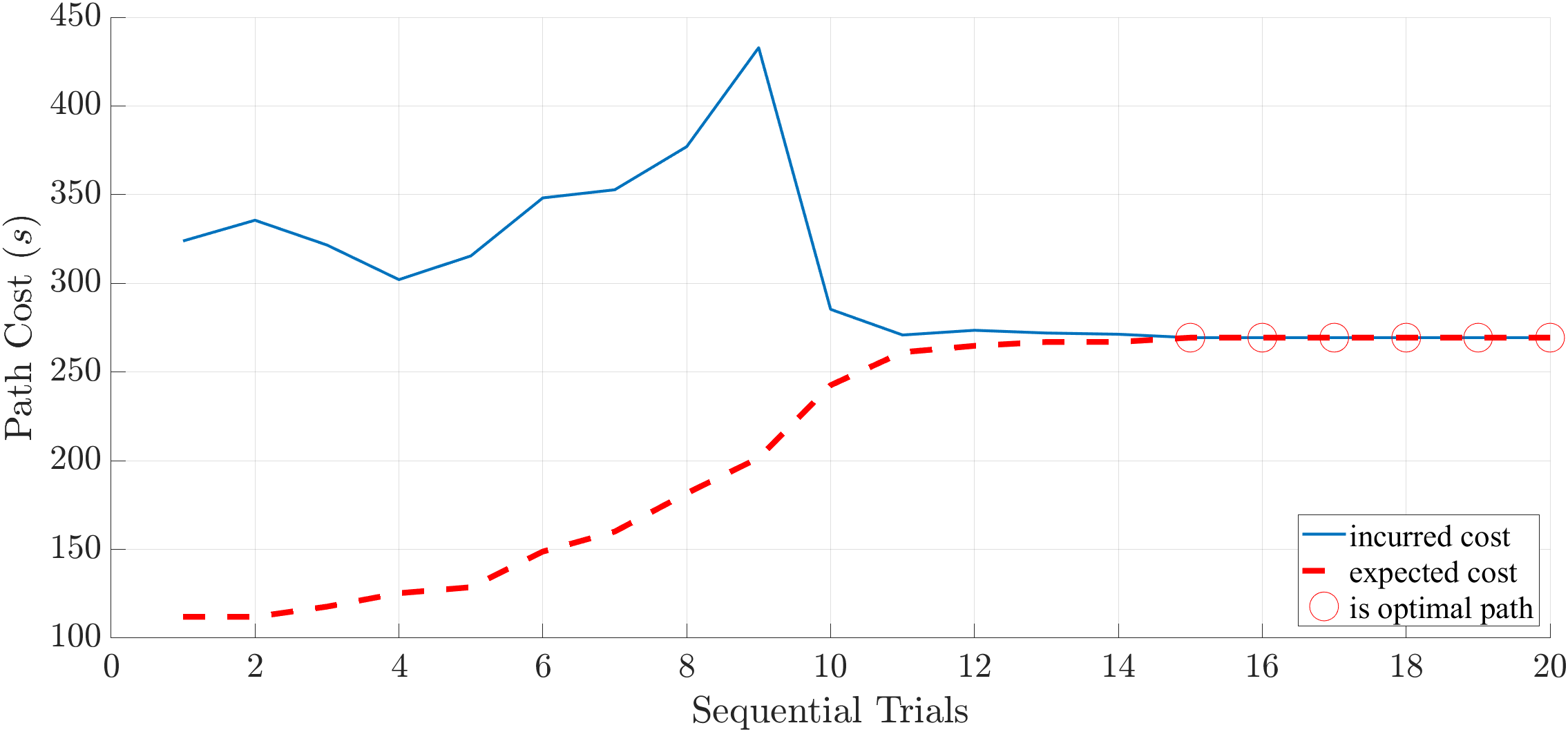}
	\caption{Expected vs incurred path cost for a $9 \times 9$ grid, environment Case 1.}
	\label{fig:9x9expectedvincurredcase1}
\end{figure}

\tbl{table_results} shows the number of passes until the optimal 
path was found, i.e., no other paths of lower cost were found in subsequent passes. 
\figfser{fig:9x9expectedvincurredcase1}{fig:9x9expectedvincurredcase3} show an example
of the evolution of the incurred and expected path costs for each pass
for Cases~1 and~3 respectively on a $9\times 9$ grid. 
The incurred cost is higher in the first few passes because the edge cost
estimates are initialized optimistically, as discussed in \scn{sec-method}.

When the true optimal path is taken, the expected data point is marked with a circle. 
In these examples, the optimal path is found after the $15\msup{th}$ pass 
and no other path is selected for future passes. Therefore, 
15 is recorded in \tbl{table_results}.

For Case 1, the expected and incurred costs are equal once the 
optimal path is reached. This is not true for Cases 2-4
due to measurement noise and estimation errors.
The number of passes required for Case~1 (no measurement noise and
constant $\dpx{2}$) is a lower bound 
for the number of passes required for any of the other cases.
\tbl{table_results} demonstrates that the proposed wind speed 
and edge cost estimation method converges to the optimal path
with the minimum or near-minimum number of passes for Cases~2--4 as well,
regardless of grid size.

\begin{figure}
	\centering
	\includegraphics[width=\columnwidth]{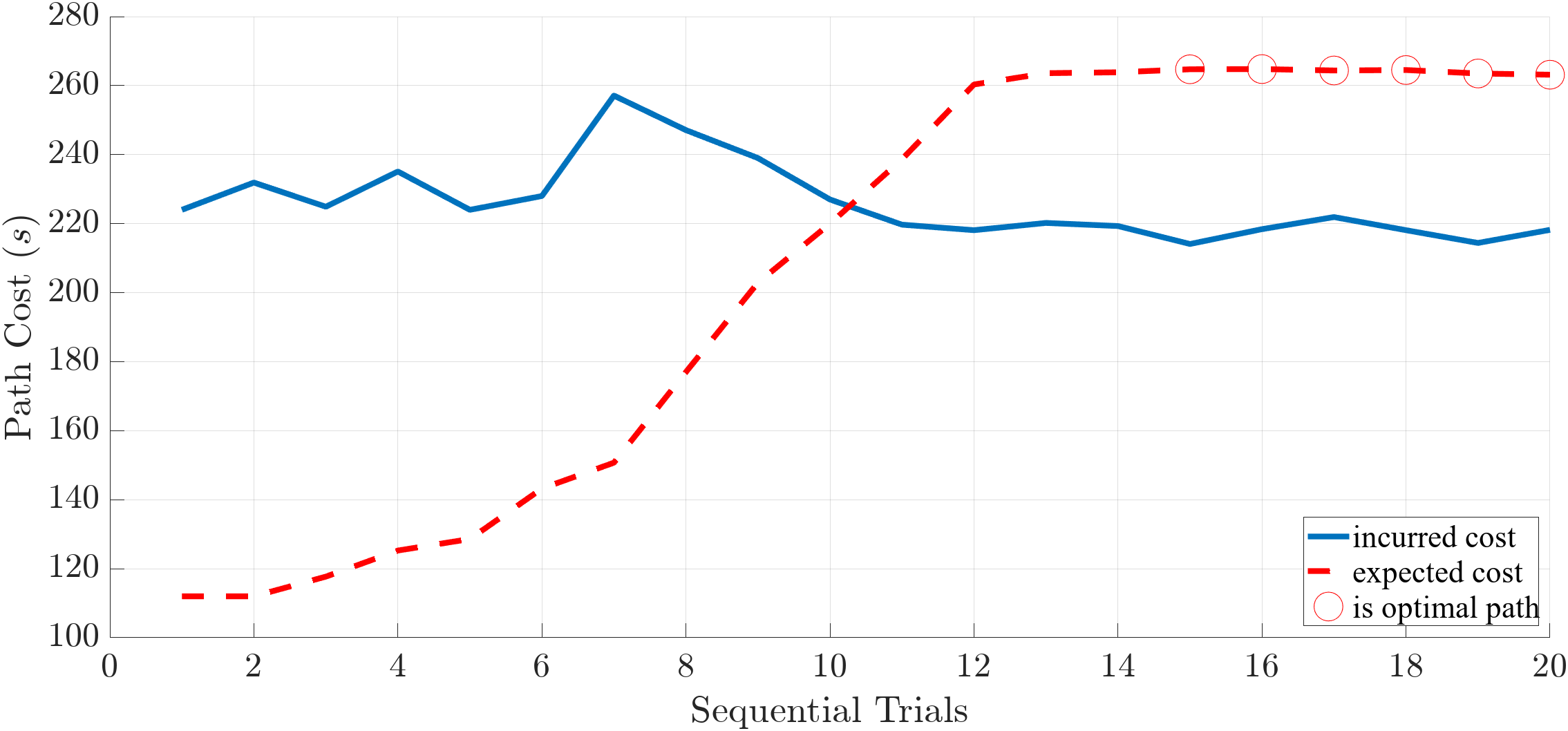}
	\caption{Expected vs incurred path cost for a $9 \times 9$ grid, environment Case 3.}
	\label{fig:9x9expectedvincurredcase3}
\end{figure}

\begin{figure}
	\centering
	\includegraphics[width=\columnwidth]{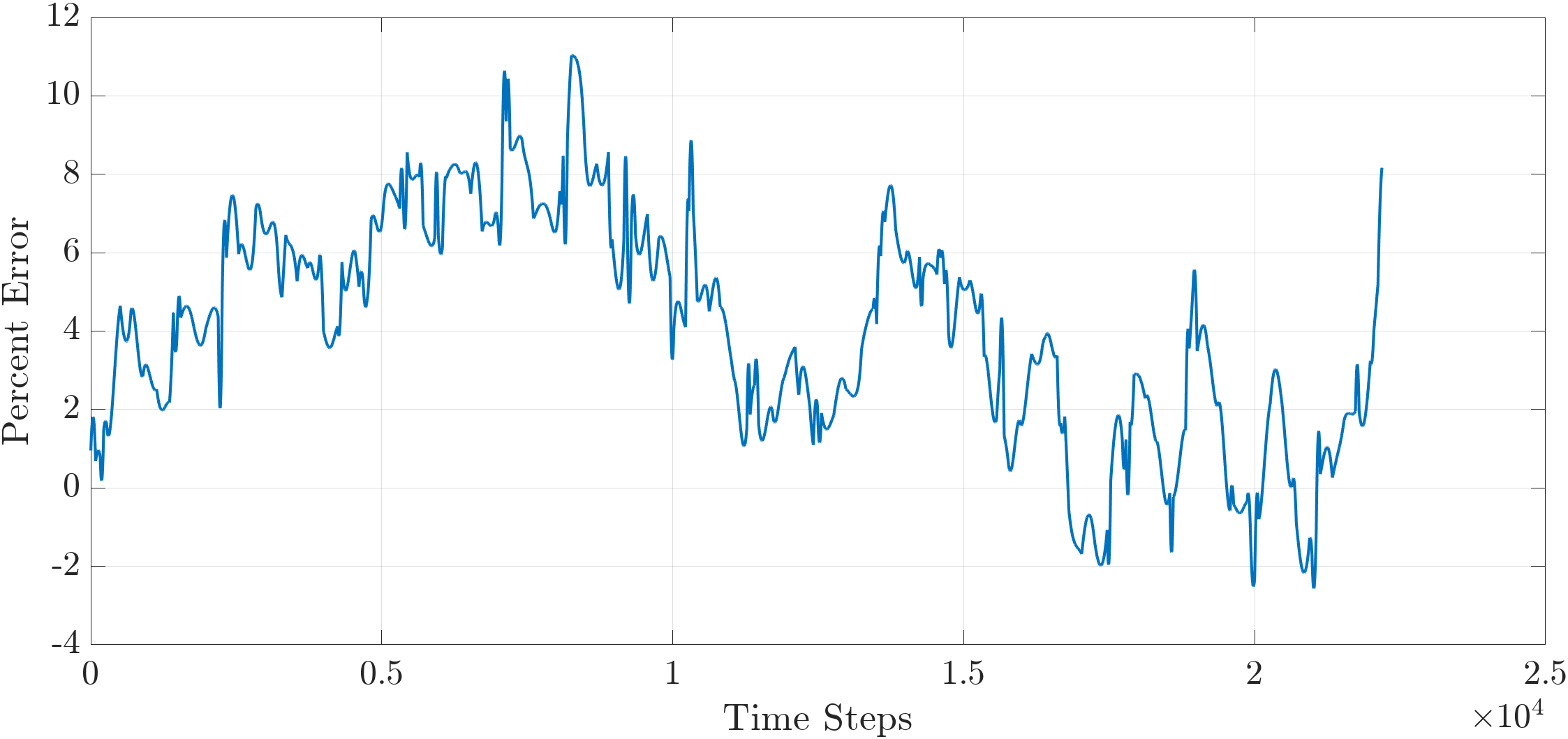}
	\caption{Percent error between the estimated and true $\dpx{2}(t)$ 
		for a $5\times 5$ grid, Case 4 without a Kalman Filter.}
	\label{fig:error5x5case4}
\end{figure}

\begin{figure}
	\centering
	\includegraphics[width=\columnwidth]{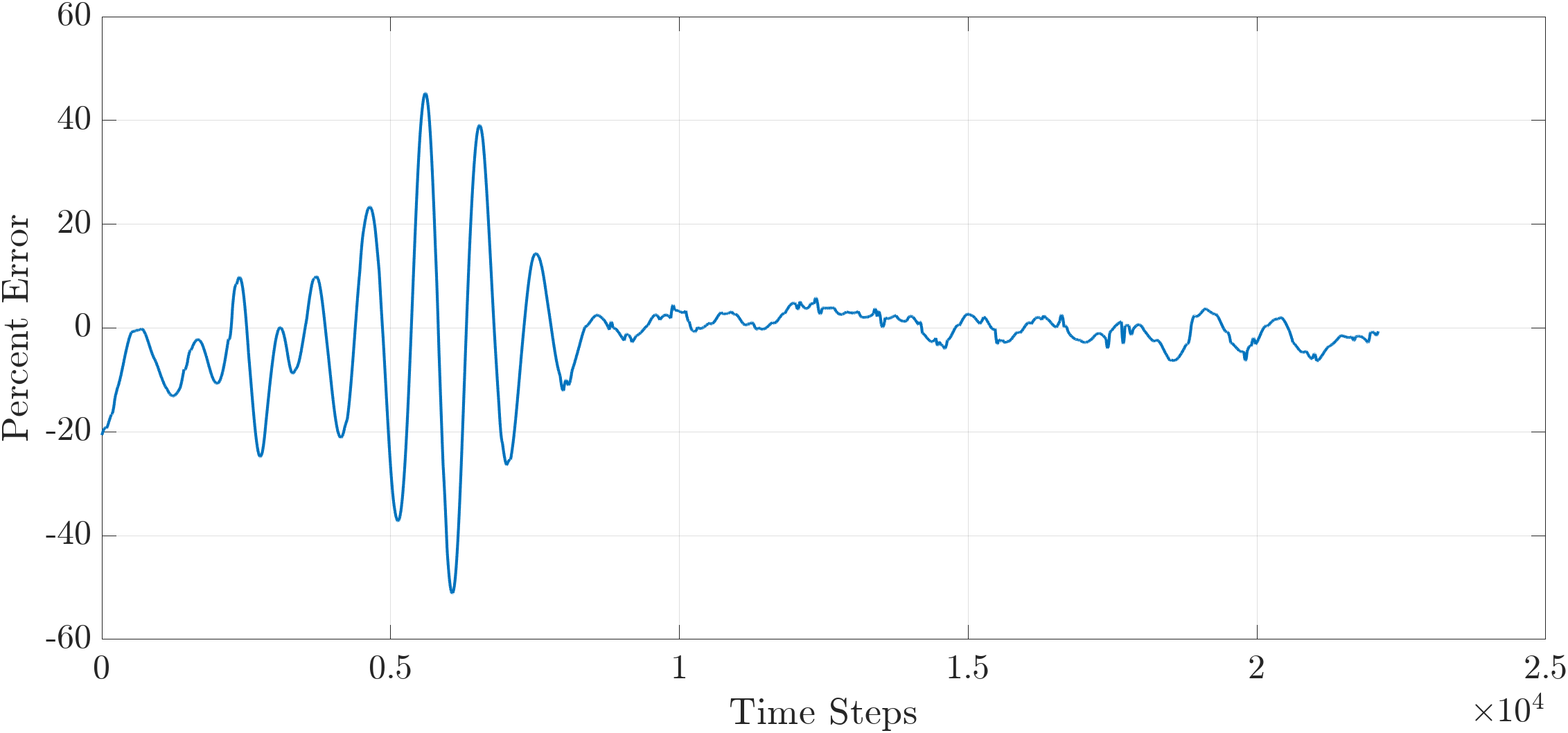}
	\caption{Percent error between the estimated and true 
		$\dpx{2}(t)$ for a $5 \times 5$ grid, Case 4 with a Kalman Filter.}
	\label{fig:error5x5case3KF}
\end{figure}

For Cases~2 and~4, the number of passes required to find the optimal path depends on
the signal to noise ratio in $\dpx{2}(t)$. For Case 4, the problem of error prorogation 
due to the polynomial curve fit (see \scn{sec-method}) becomes prominent.
\figfser{fig:error5x5case4}{fig:error5x5case3KF} show that the steady state error
can be reduced by applying a Kalman Filter to estimate $\dpx{2}(t)$.

The Kalman Filter can suffer from high 
errors early in the signal while the input data window is too small for the FFT.
This does not alter performance 
since $\edgeCost{i}{j}$ is determined using the most recent measurements
$\windMeas{i}{j}(n:m)$ available, which are then used to estimate 
$\edgeResist{i}{j}$ by \eqnnt{eq-resistance-calc}.

Once the optimal path is found, the Kalman Filter better 
approximates $\dpx{2} (n:m)$ as $n$ gets large
since it is less prone to propagating errors. However, using
the Kalman filter did not lead to the optimal path in
fewer passes than the polynomial curve fit-based estimation alone.

\section{Conclusions}

\label{sec-conclusions}

We studied the problem of planning the paths of a team of SUAVs
sequentially flying from a fixed initial point to a fixed destination
in an unknown wind field. The novel aspect of our problem formulation
is that each SUAV measures wind speeds while traversing along a 
minimum-cost path planned based on the current wind field estimate.
We provided a technique of estimating the wind field such that the
optimal (minimum-time) path is found after a small number of passes.
We considered cases of static and periodically time-varying winds 
with and without measurement noise. The number of passes required 
for the easiest case (static wind, no measurement noise) is the 
minimum required for all other cases. We demonstrated numerical 
simulation results where optimal paths were found for all other cases 
with a minimum or near-minimum number of passes.
Future work will expand the problem to multiple SUAVs running concurrently
and a simulation with high-fidelity flow fields and SUAV flight dynamics.

\addtolength{\textheight}{-11cm}   % This command serves to balance the column lengths

\bibliographystyle{ieeetr}
\bibliography{Reference}

\begin{thebibliography}{10}

\bibitem{Cohen2021}
A.~P. Cohen, S.~A. Shaheen, and E.~M. Farrar, ``Urban air mobility: History,
  ecosystem, market potential, and challenges,'' {\em IEEE Transactions on
  Intelligent Transportation Systems}, vol.~22, no.~9, pp.~6074--6087, 2021.

\bibitem{Zhao2024}
B.~Zhao, Y.~Suo, L.~Tang, C.~Li, M.~Fu, and L.~Huang, ``Urban air mobility for
  time-sensitive goods with explicit customer preferences: A case study on
  chengdu,'' {\em Journal of Air Transport Management}, vol.~118, p.~102613,
  2024.

\bibitem{goyal2018urban}
R.~Goyal, C.~Reiche, C.~Fernando, J.~Serrao, S.~Kimmel, A.~Cohen, and
  S.~Shaheen, ``Urban air mobility (uam) market study,'' Tech. Rep.
  HQ-E-DAA-TN65181, 2018.
\newblock Available: \url{https://ntrs.nasa.gov/citations/20190001472}.

\bibitem{Rakib2020}
M.~Rakib, S.~Evans, and P.~Clausen, ``Measured gust events in the urban
  environment, a comparison with the iec standard,'' {\em Renewable Energy},
  vol.~146, pp.~1134--1142, 2020.

\bibitem{CHRIT2023100451}
M.~Chrit, ``Reconstructing urban wind flows for urban air mobility using
  reduced-order data assimilation,'' {\em Theoretical and Applied Mechanics
  Letters}, vol.~13, no.~4, p.~100451, 2023.

\bibitem{doi:10.2514/6.2020-2890}
M.~Xue and M.~Wei, {\em Small UAV Flight Planning in Urban Environments}.

\bibitem{WU2024105253}
J.~Wu, Y.~Ye, and J.~Du, ``Multi-objective reinforcement learning for
  autonomous drone navigation in urban areas with wind zones,'' {\em Automation
  in Construction}, vol.~158, p.~105253, 2024.

\bibitem{refId0}
{Korprasertsak, Natapol} and {Leephakpreeda, Thananchai}, ``Improving accuracy
  of wind analysis with multiple sampling rates of wind measurement,'' {\em E3S
  Web Conf.}, vol.~95, p.~02002, 2019.

\bibitem{LaValle2006}
S.~M. Lavalle, {\em Planning Algorithms}.
\newblock Cambridge University Press, 2006.

\bibitem{Patle2019}
B.~K. Patle, G.~B. L, A.~Pandey, D.~R. Parhi, and A.~Jagadeesh, ``A review: On
  path planning strategies for navigation of mobile robot,'' {\em Defence
  Technology}, vol.~15, pp.~582--606, August 2019.

\bibitem{wen2024drl}
T.~Wen, X.~Wang, Z.~Zheng, and Z.~Sun, ``A \uppercase{DRL}-based path planning
  method for wheeled mobile robots in unknown environments,'' {\em Computers
  and Electrical Engineering}, vol.~118, p.~109425, 2024.

\bibitem{qin2023deep}
Y.~Qin, Z.~Zhang, X.~Li, W.~Huangfu, and H.~Zhang, ``Deep reinforcement
  learning based resource allocation and trajectory planning in integrated
  sensing and communications \uppercase{UAV} network,'' {\em IEEE Transactions
  on Wireless Communications}, vol.~22, no.~11, pp.~8158--8169, 2023.

\bibitem{Lewis2017optimal}
F.~L. Lewis, L.~Xie, and D.~Popa, {\em Optimal and robust estimation: with an
  introduction to stochastic control theory}.
\newblock CRC press, 2017.

\bibitem{myung2003tutorial}
I.~J. Myung, ``Tutorial on maximum likelihood estimation,'' {\em Journal of
  mathematical Psychology}, vol.~47, no.~1, pp.~90--100, 2003.

\bibitem{thrun2006}
S.~Thrun, W.~Burgard, and D.~Fox, {\em Probabilistic Robotics}.
\newblock The MIT Press, 2006.

\bibitem{Julier2004}
S.~J. Julier and J.~K. Uhlmann, ``Unscented filtering and nonlinear
  estimation,'' in {\em Proceedings of the IEEE}, vol.~92, pp.~401--422, Mar.
  2004.

\bibitem{deng2021poserbpf}
X.~Deng, A.~Mousavian, Y.~Xiang, F.~Xia, T.~Bretl, and D.~Fox, ``Poserbpf: A
  rao--blackwellized particle filter for 6-d object pose tracking,'' {\em IEEE
  Transactions on Robotics}, vol.~37, no.~5, pp.~1328--1342, 2021.

\bibitem{HEIJ1999993}
C.~Heij and W.~Scherrer, ``Consistency of system identification by global total
  least squares,'' {\em Automatica}, vol.~35, no.~6, pp.~993--1008, 1999.

\bibitem{LOUARROUDI201113115}
E.~Louarroudi, J.~Lataire, and R.~Pintelon, ``Frequency domain total least
  squares identification of linear, periodically time-varying systems from
  noisy input-output data,'' {\em IFAC Proceedings Volumes}, vol.~44, no.~1,
  pp.~13115--13120, 2011.
\newblock 18th IFAC World Congress.

\bibitem{LIU2020736}
G.~Liu, M.~li~Yu, L.~Wang, Z.~yi~Yin, J.~ke~Liu, and Z.~rong Lu, ``Rapid
  parameter identification of linear time-delay system from noisy frequency
  domain data,'' {\em Applied Mathematical Modelling}, vol.~83, pp.~736--753,
  2020.

\bibitem{poudel2024coupled}
P.~Poudel and R.~V. Cowlagi, ``Coupled sensor configuration and planning in
  unknown dynamic environments with context-relevant mutual information-based
  sensor placement,'' in {\em 2024 American Control Conference (ACC)},
  pp.~306--311, IEEE, 2024.

\bibitem{Poudel-Cowlagi-Scitech2025}
P.~Poudel and R.~V. Cowlagi, ``Reconfiguration costs in coupled sensor
  configuration and path-planning for dynamic environments,'' in {\em AIAA
  Scitech 2025 Forum}, (Orlando, Florida), 6-10 January 2025.
\newblock to appear.

\bibitem{sarras2014guidance}
I.~Sarras and H.~Siguerdidjane, ``On the guidance of a uav under unknown wind
  disturbances,'' in {\em 2014 IEEE Conference on Control Applications (CCA)},
  pp.~820--825, IEEE, 2014.

\bibitem{Zhou2017}
B.~Zhou, H.~Satyavada, and S.~Baldi, ``Adaptive path following for unmanned
  aerial vehicles in time-varying unknown wind environments,'' in {\em 2017
  American Control Conference (ACC)}, pp.~1127--1132, 2017.

\bibitem{Yang2021}
J.~Yang, C.~Liu, M.~Coombes, Y.~Yan, and W.-H. Chen, ``Optimal path following
  for small fixed-wing uavs under wind disturbances,'' {\em IEEE Transactions
  on Control Systems Technology}, vol.~29, no.~3, pp.~996--1008, 2021.

\bibitem{Xing2023}
Z.~Xing, Y.~Zhang, and C.-Y. Su, ``Active wind rejection control for a
  quadrotor uav against unknown winds,'' {\em IEEE Transactions on Aerospace
  and Electronic Systems}, vol.~59, no.~6, pp.~8956--8968, 2023.

\bibitem{liu2016disturbance}
C.~Liu and W.-H. Chen, ``Disturbance rejection flight control for small
  fixed-wing unmanned aerial vehicles,'' {\em Journal of Guidance, Control, and
  Dynamics}, vol.~39, no.~12, pp.~2810--2819, 2016.

\bibitem{Wang2019}
B.~H. Wang, D.~B. Wang, Z.~A. Ali, B.~T. Ting, and H.~Wang, ``An overview of
  various kinds of wind effects on unmanned aerial vehicle,'' {\em Measurement
  and Control}, vol.~52, no.~7-8, pp.~731--739, 2019.

\bibitem{Cacan2015}
M.~R. Cacan, E.~Scheuermann, M.~Ward, M.~Costello, and N.~Slegers, ``Autonomous
  airdrop systems employing ground wind measurements for improved landing
  accuracy,'' {\em IEEE/ASME Transactions on Mechatronics}, vol.~20, no.~6,
  pp.~3060--3070, 2015.

\bibitem{cecil2017hurricane}
D.~J. Cecil and S.~K. Biswas, ``Hurricane imaging radiometer (hirad) wind speed
  retrievals and validation using dropsondes,'' {\em Journal of atmospheric and
  oceanic technology}, vol.~34, no.~8, pp.~1837--1851, 2017.

\end{thebibliography}

\end{document}